\documentclass[prd,preprintnumbers,nofootinbib,aps]{revtex4}

\usepackage{epsfig,amsmath,amssymb}
\usepackage{mathrsfs}
\usepackage[usenames,dvipsnames]{color}
\usepackage[pagebackref=false, colorlinks=false]{hyperref}

\begin{document}

\title{Kinematics of two-particle scattering in black hole backgrounds}

\author{Soumendra Kishore Roy}
\email{soumendrakishoreroy09@gmail.com} 
\affiliation{Department of Physics, Presidency University, Kolkata 700073, 
India.}

\author{Ratna Koley}
\email{ratna.physics@presiuniv.ac.in} 
\affiliation{Department of Physics, Presidency University, Kolkata 700073, 
India.}  

\author{Parthasarathi Majumdar}
\email{bhpartha@gmail.com}
\affiliation{School of Physical Sciences, Indian Association for the Cultivation of Science, Kolkata 700032, India }

\begin{abstract}

We show that particle scattering in general curved backgrounds entails {\it six} independent, kinematical Mandelstam-like invariants, instead of the 
two in flat spacetime. Spacetime isometries are shown to lead to constraints between these parameters, so that for standard black holes like Schwarzschild, 
Reissner-Nordstr\"om, or Kerr spacetimes, the number of {\it independent} parameters may be less than six. We compute the values of these independent 
parameters very close to the event horizon of the black holes. We demonstrate the existence of kinematical domains in the parameter space of particle 
trajectories for which some of the independent invariants may become unbounded above, as the point of collision approaches the event horizon. 
For particle scattering, this would imply the possibility of scattering with very large center-of-mass energy squared and/or very large 
momentum-transferred squared, making this astrophysically a laboratory for physics beyond the standard model.

\end{abstract}

\maketitle

\section{Introduction}

The center-of-mass energy for a pair of particles
falling towards a Kerr or Reissner-Nordstr\"om (RN) black hole can become unbounded from above at the horizon
(\cite{BSW}, \cite{RN}, \cite{Kerr-Newman}, and \cite{Kerr-Newman1}) for
specific ranges of angular momenta or electric charges of the incident
particles.
This unboundedness is independent of the Thorne bound \cite{thr}, \cite{thr1}
(a bound on the energy of the particle per unit Kerr parameter \cite{thorne}). These effects hold true for various astrophysical scenarios like in the presence of the magnetic field 
\cite{MagneticField}, \cite{MagneticField1} and in the context of observability
of high energy particles in astrophysical black hole backgrounds
\cite{Astro}-\cite{Astro4}. It is by now established that the
effect of the unboundedness in the center-of-mass energy in certain
kinematical domains are a universal property of classical particle
kinematics in the background of asymptotically flat black holes
\cite{RN}, \cite{Universal}. The issue of the divergence of the center of mass-energy is also extended to the collision of the two particles for stringy \cite{Sen} black holes and higher dimensional Kerr black hole backgrounds \cite{MP}, \cite{MP1}.

In this paper, we consider the prospects of black holes as particle accelerators from a more general standpoint. The kinematics of two-particle scattering in flat spacetime is characterized by two independent parameters from a linearly dependent triplet known as the Mandelstam parameters, written as $s~,~t$, and $u$. Of these, $s$ is the total squared center-of-mass energy, and it is the analog of this quantity in curved black hole spacetime whose unboundedness from above in certain kinematical domains has been discussed extensively in the literature. The question that naturally arises is: what about the other Mandelstam parameters $t$ and $u$ which are essential for a complete description of particle scattering in flat spacetime? Do they or do they not diverge in the same kinematical regimes in black hole backgrounds. Or maybe in other regimes? But preceding this query is the question that in the curved black hole background, are the possible Mandelstam-like scalars, parametrizing the kinematics of two-particle scattering, only the ones dealt with in flat spacetime, or are there others? If the latter, then the issue of the unboundedness or otherwise of the other invariants in kinematical regimes does indeed arise.

We show that in general there are {\it six} independent kinematical invariants that need to be investigated,
instead of the center-of-mass energy alone. The isometries of the asymptotically flat black hole spacetimes give another two constraint relations among the six Mandelstam-like invariants. We limit our study for the two-particle collision in the equatorial plane of the axissymmetric stationary black hole spacetime for which the number of kinematical invariants is only {\it three}. We calculate these invariants for both particle-pairs propagating along geodesics as well as those which follow nongeodesic paths, for various
asymptotically flat black hole backgrounds. We find that a class of the six invariants can become unbounded in certain kinematical regimes.

In this paper we ignore the backreaction of the particle energies on the black hole geometry. In other words, irrespective of the values of the kinematical parameters, the background metric  is assumed to remain the same. The reason behind this assumption is that the change in geometry due to a localized object  like a particle is in the nature of a small gravitational shock wave passing any location in the spacetime momentarily.  So the scattering of two particles cannot cause any major change in the overall curvature profile of a stellar or a supermassive black hole. However, the backreaction may be of importance if one is dealing with a {\it beam} of very highly energetic particles, scattering off another similar beam.  We also neglect the effect of the gravitational and electromagnetic fields on any one of the particles due to the other particle.

The paper is structured as follows : Sec: II reviews the flat spacetime arguments underlying the Mandelstam parameters. It is shown how two-particle scattering in an arbitrary curved spacetime must entail a larger set of kinematical invariants. In the next section, these invariants are computed for the scattering of two neutral charged for spherically symmetric black hole spacetimes (Schwarzschid and Reissner-Nordstr\"om), and their possible unboundedness demonstrated to the existence of a sharp asymmetry between the geodesically conserved energies of the particles. Sec: IV considers charged particle scattering in a Reissner-Nordstr\"om black hole background, and the possibility of unboundedness of the kinematical invariants is shown to occur, for generic energies, angular momenta and charges, when a certain condition is imposed relating the parameters.  This exercise is repeated for particle scattering in Kerr black hole backgrounds in Sec: V, and charged particle scattering in Kerr-Newman black hole backgrounds in Sec: VI, leading to a general criterion for the possible unboundedness of the kinematical invariants in particle scattering very close to the horizon of black hole spacetimes. We conclude in Sec VII.

\section{Independent invariants for two-particle scattering in curved spacetime}

Let us consider 2-to-2 particles scattering in a general curved spacetime. The 
kinematical properties of this collision are 
characterized by the four momenta of incoming and outgoing particles. {\color{Brown} Thus} the scattering cross section $\sigma_{1,2 \rightarrow 3,4}$ is a function of the four $4$ momenta, {\it{i.e.}} of 
$sixteen$  variables. The cross section, $\sigma_{1,2 
\rightarrow 3,4}$, should be a frame invariant quantity; therefore, 
it must  be a function of scalar products of the four momenta. Hence, the 
number of variables to characterize collision depends on $ten$
scalars in four-dimensional spacetime. 

Neglecting the backreaction, the action for a particle of rest mass $m$ 
in a spacetime with metric $g_{\mu \nu}$ is given by, in the units $c=1$,
\begin{equation}\label{action}
S=(-m)\int_P^Q\sqrt{g_{\mu \nu}dx^{\mu}dx^{\nu}}=(-m)\int_{\tau_P}^{\tau_Q}\sqrt{g_{\mu \nu}u^{\mu}u^{\nu}}d\tau
\end{equation}
where $u^{\mu}$ is the four velocity and the particle goes from P to Q in four-dimensional spacetime with proper time $\tau$.

Let us denote four vectors by boldface capital letters and components by small-case letters with greek superscripts and subscripts. The canonical four momentum corresponding to \eqref{action} is given by, 
\begin{equation}\label{defn}
p^{\mu}=mu^{\mu}
\end{equation}
so that, the energy-momentum dispersion relation for a free particle in curved spacetime is 
\begin{equation}\label{mass-hyperboloid}
\textbf{p}^2=p_{\mu}p^{\mu}=m^2g_{\mu \nu}u^{\mu}u^{\nu}=m^2 
\end{equation}
as, $g_{\mu \nu}u^{\mu}u^{\nu}=1$ for a timelike geodesic particle in the natural unit system. Equation \eqref{mass-hyperboloid} is valid for each of the four particles, so it gives further four scalar constraints on the cross section $\sigma_{1,2 \rightarrow 3,4}$. The number of independent kinematic invariants thus reduces and becomes {\it six.}

Until this point, the argument for the number of independent invariants is the same as in Minkowski spacetime. However, in Minkowski spacetime, one can invoke spacetime translational symmetry corresponding to an arbitrary constant shift of the spacetime origin. The conserved N\"other charges corresponding to this symmetry are the {\it total} four momenta ${\bf p}_{tot} \Rightarrow {\bf p}_1 +{\bf p}_2= {\bf p}_3 + {\bf p}_4$. This corresponds to four additional constraints which cut down the number of independent invariants to characterize two-particle scattering in flat spacetime to {\it two}.
 
However, such a global spacetime translational symmetry in full generality is absent in a curved spacetime. So the additional argument to reduce the number of independent invariants characterizing two-particle scattering to two, from six, is missing. Thus, in a general curved spacetime, we are left with the six invariants mentioned above. Depending on the isometries of a particular spacetime, there may be additional conserved quantities corresponding to geodesic propagation, reducing thus the number of independent invariants further. Hence $\textbf{p}_{(1)}+\textbf{p}_{(2)} \neq \textbf{p}_{(3)}+\textbf{p}_{(4)}$, and the number of independent invariants on which the scattering cross section depends is in general six.  

Let us define these six parameters as
\begin{equation}\label{eq:Mandelstam}
s_{ij}=(\textbf{p}_{(i)}+\eta~\textbf{p}_{(j)})^2~i,j=1,2,3,4;~i<j
\end{equation}
In the channel where 1,2 are colliding and 3,4 are being produced, $\eta=+1$ for $i=1$, $j=2$ and $i=3$, $j=4$\\
and $=-1$ otherwise.

We say these independent parameters are six Mandelstam-like scalars as in flat spacetime $s_{12}$, $s_{13}$, and $s_{14}$ become equals to $s_{34}$, $s_{24}$, and $s_{23}$ and gives Mandelstam variables $s$, $t$, and $u$ respectively \cite{Book}. In an arbitrarily curved spacetime, $s_{12}$ and $s_{34}$ give the center of mass-energy in the center of mass frame of $1,2$ and $3,4$, respectively (Due to lack of 4-momentum conservation, these two frames are different, and hence, these two invariants are not identical). Similarly, $s_{13}$ and $s_{14}$ refer to the momentum transfer and cross-momentum transfer in the CM frame of 1, 2, and $s_{24}$ and $s_{23}$ give the momentum transfer and cross-momentum transfer in the CM frame of 3,4.

The physical motivation for considering such generalized kinematical invariants in particle scattering in black hole spacetimes stems from the deeper underlying possibility of particle scattering involving large center-of-mass energy {\it together with} large momentum transfer where short-range effects due to standard strong-electroweak theory and beyond Standard Model interactions may begin to manifest. This would lend greater credence to the idea of black hole spacetimes as particle accelerators. Despite the localized nature of the kinematical invariants in a curved spacetime implying various restrictions \cite{Astro}-\cite{Astro4} on the measurement of these invariants away from the horizon, local particle interactions with large values of the invariants {\it close} to the interaction point may still lead to new physics. Thus, in general, all {\it six} independent kinematical invariants need to be investigated.

For spacetime with isometries (like asymptotically flat Schwarzschild, Reissner Nordstr\"om or Kerr black hole), additional constraint relations are imposed, and for the equatorial two-particle collision, the number of independent kinematical parameters are found to be {\it three}. In the next stage, we explicitly see the mentioned reduction and calculate the relevant kinematical invariants for several asymptotically flat black hole spacetimes. We show indeed some cases lead to kinematical unboundedness.

\section{Neutral Particle Scattering in the Spherical Symmetric Background}

The metric of any spherically symmetric spacetime in the metric convention (+1,-1,-1,-1) is given by
\begin{equation}\label{metric}
ds^2=f(r)dt^2-g(r)dr^2-r^2(d\theta^2+\sin^2\theta d\phi^2)
\end{equation}
where $f(r)$ and $g(r)$ are positive functions. The four velocity of a particle in the equatorial plane of the metric (\ref{metric}) is found by integrating the geodesic equation once,
\begin{equation}
\frac{dt}{d\tau}=f^{-1}E,~\frac{dr}{d\tau}=\sqrt{f^{-1}g^{-1}E^2-\bigg(1+\frac{L^2}{r^2}\bigg)g^{-1}},~\theta=\pi/2,~\frac{d\phi}{d\tau}=\frac{L}{r^2}
\end{equation}
where E and L are the geodesic constants energy and angular momentum per unit rest mass of the particle. So the four-momentum of the $i^{th}$ particle is
\begin{equation}\label{eq:4P1}
\textbf{P}_{(i)}~=~m(u^t_{(i)},~ u^r_{(i)},~ u^{\theta}_{(i)},~ u^{\phi}_{(i)})~=~m\bigg(f^{-1}E_{(i)},~\sqrt{f^{-1}g^{-1}E^2_{(i)}-\bigg(1+\frac{L^2_{(i)}}{r^2}\bigg)g^{-1}},~0,~\frac{L_{(i)}}{r^2}\bigg)~~i=1,2,3,4
\end{equation}
where we assume the produced particles to follow timelike geodesics and the rest masses of all four particles are the same.

The metric (\ref{metric}), allows two Killing vectors corresponding to $t$ and $\phi$, which leads to conservation of $E$ and $L$ $\Rightarrow$ $E_{(1)}+E_{(2)}=E_{(3)}+E_{(4)}$ and $L_{(1)}+L_{(2)}=L_{(3)}+L_{(4)}$. These two relations give another two constraints on the six Mandelstam-like invariants and reduces the number of independent invariants to {\it four} for this particular background spacetime. We are also considering a collision in a particular plane (precisely, the equatorial plane), and thus, the $p^{\theta}$ component of the four momenta of the particles is always conserved ($p^{\theta}_{(1)}+p^{\theta}_{(2)}=p^{\theta}_{(3)}+p^{\theta}_{(4)}$). Hence, for this particular kind of collision, the number of independent Mandelstam-like invariants comes out to be {\it three}. We evaluate all of the six invariants here keeping in mind that only three of them are actually independent.

For a point collision, the metric and all of the four momenta of four particles are a function of the same radial coordinate $r$ and the six Mandelstam-like scalars are given for the particles of same mass $m$,
\begin{center}
$s_{ij}=( \textbf{p}_{(i)}~+~\eta \textbf{p}_{(j)})^2=\textbf{p}_{(i)}^2+\textbf{p}_{(j)}^2+2~\eta~ \textbf{p}_{(i)}.\textbf{p}_{(j)}=2m^2(1+\eta ~ g_{\mu \nu}~u^{\mu}_{(i)}~u^{\nu}_{(j)})$
\end{center}
as by \eqref{mass-hyperboloid}, $\textbf {p}_{(i)}^2=m^2$. For the spherically symmetric metric given in \eqref{metric}, the six scalars at the radial coordinate $r$ are as follows:
\begin{equation}
\frac{s_{ij}}{2m^2}~=~\frac{fr^2+\eta ~ [E_{(i)}E_{(j)}r^2-\sqrt{E_{(i)}^2r^2-fr^2-L_{(i)}^2f}~\sqrt{E_{(j)}^2r^2-fr^2-L_{(j)}^2f}-L_{(i)}L_{(j)}f]}{fr^2} \label{6scalars1}
\end{equation}
For a spherical distribution of matter with a diagonal $T_{\mu \nu}$ satisfying $T_{tt}=T_{rr}$ (it can be zero also) gives
\begin{equation}\label{fg}
g~=~f^{-1}
\end{equation}
For \eqref{fg}, the spacetime allows a black hole horizon {\it i.e.}, for some non zero $r,~f(r)$ is zero. For some cases, the metric allows two horizons spacetime also, but we restrict ourselves only to the outer horizon. In this case, the expression is indeterminate, and so we take the appropriate limit.

At the horizon of a nonextremal black hole, $f=0$, but $f^{\prime} \neq 0$. The limit $f \rightarrow 0$ with $f ^{\prime} \neq 0$ gives
\begin{equation}\label{shorizon}
\frac{s_{ij}}{2m^2}|_{Horizon}~=~1+\frac{\eta}{2} \Bigg[\frac{E_{(i)}}{E_{(j)}}\Bigg(1+\frac{L_{(j)}^2}{r_H^2}\Bigg)+\frac{E_{(j)}}{E_{(i)}}\Bigg(1+\frac{L_{(i)}^2}{r_H^2}\Bigg)-2\frac{L_{(i)}L_{(j)}}{r_H^2}\Bigg]
\end{equation}
where, $r=r_H$ gives the horizon, i.e. $f(r_H)=0$. For the Schwarzschild and Reissner-Nordstr\"om black holes $r_H$ is given by $2M$ and $M+\sqrt{M^2-Q^2}$, respectively ($M$ is the mass of both black holes and $Q$ is the charge of the Reissner-Nordstr\"om black hole).

For an extremal black hole, both $f$ and $f^{\prime}$ appearing in the metric \eqref{metric} are zero at the horizon. We take the horizon limit $f \rightarrow 0$ first then put the extremality condition, $f^{\prime} \rightarrow 0$ in the expression \eqref{6scalars1}. Taking these limits, six Mandelstam-like scalars at the horizon of the spherical symmetric extremal black hole turns out to be the same as \eqref{shorizon}. Now the extremal limit is realized as $M=Q$, and the horizon is changed to $r_H=M=Q$ in the case of a Reissner Nordstr\"om black hole.

The quantity in \eqref{shorizon} can indeed be made very large in the kinematical domain where $|E_{(i)}|>>|E_{(j)}|$, or vice versa, {\it i.e.} under the condition that one particle has a far higher energy compared to the other at asymptotpia within an asymptotically flat spacetime. It is the energy {\it asymmetry} between the particles in the far past, which might lead to very large kinematical invariants near the horizon.

For the particles colliding with their rest energy ($m$) and angular momenta $L_{(1)}=4GM$, $L_{(2)}=-4GM$ at the horizon of the Schwarzschild black hole with a mass $M=1$, $s_{12}$ or a center of mass energy squared comes out to be $20m^2$, which is in agreement with \cite{BSW}.

\section{Charged Particles in the Spherical Symmetric spacetime}

The Eq. \eqref{metric} above gives the general metric of the spherically symmetric spacetime in the $(t,r,\theta,\phi)$ basis. The particles being charged experience a pull by the electromagnetic field apart from the gravitation. The action for a charged particle of mass $m$ and charge $e$ in the spacetime with a metric $g_{\mu \nu}$ and four potential $A^{\mu}$ is (in the unit $c=1$), 
\begin{equation}\label{Action2}
S=(-m)\int^B _A\sqrt{g_{\mu \nu}dx^{\mu}dx^{\nu}}-e\int^B_A g_{\mu \nu}A^{\mu}dx^{\nu}
\end{equation}
Extremization of the \eqref{Action2} gives,
\begin{equation}\label{charged}
\frac{du^{\mu}}{d\tau}+\Gamma^{\mu}_{\nu \lambda}u^{\nu}u^{\lambda}=\frac{e}{m}F^{\mu}_{\nu}u^{\nu}
\end{equation}
Equation \eqref{charged} can be integrated for a spacetime having a metric (\ref{metric}) and the electromagnetic field produced by a static charge $Q$, to get the four momenta of the timelike test charge moving in the equatorial plane, 
\begin{equation}\label{4p}
\frac{dt}{d\tau}=f^{-1}(E-\frac{eQ}{r}),~\frac{dr}{d\tau}=\sqrt{f^{-1}g^{-1}(E-\frac{eQ}{r})^2-(1+\frac{L^2}{r^2})g^{-1}},~\frac{d\theta}{d\tau}=0,~\frac{d\phi}{d\tau}=\frac{L}{r^2}
\end{equation} 
where $E$ and $L$ are constants along the trajectory and they are interpreted as energy and angular momentum per unit test mass.

The $t$ and $\phi$ symmetry of the Lagrangian corresponding to the action (\ref{Action2}) with the metric (\ref{metric}) guarantees conservation of $E$ and $L$, {\it i.e.,} $E_{(1)}+E_{(2)}=E_{(3)}+E_{(4)}$ and $L_{(1)}+L_{(2)}=L_{(3)}+L_{(4)}$. The collision of the particles in a particular plane (equatorial plane), gives one further constraint  $p^{\theta}_{(1)}+p^{\theta}_{(2)}=p^{\theta}_{(3)}+p^{\theta}_{(4)}$. With all of these together, the number of independent Mandelstam-like invariants reduces to {\it three} like the case of neutral particle scattering. The charge conservation is assumed to be valid here, so it does not affect the number of independent parameters.

The six scalars (among which only $three$ are independent) for the particles of the same mass $m$ are calculated to be-
\begin{equation}\label{c6}
\frac{s_{ij}}{2m^2}=1+\frac{\eta}{f}\bigg[{\cal  E}_{(i)} {\cal E}_{(j)}-\sqrt{{\cal E}_{(i)}^2-f\bigg(1+\frac{L_{(i)}^2}{r^2}\bigg)}\sqrt{{\cal E}_{(j)}^2-f\bigg(1+\frac{L_{(j)}^2}{r^2}\bigg)}-\frac{L_{(i)}L_{(j)}f}{r^2}\bigg]
\end{equation}
where, ${\cal E}_{(i)}=E_{(i)}-\frac{e_{(i)}Q}{r}$.

For $T^t_t=T^r_r=\frac{Q^2}{8\pi r^4}$, \eqref{metric} allows a two horizons Reissner Nordstr\"om black hole solution having $f=g^{-1}=1-\frac{2M}{r}+\frac{Q^2}{r^2}$ and the outer horizon is located at $r_H=M+\sqrt{M^2-Q^2}$. Expression (\ref{c6}) is in the zero by zero form at $r=r_H$ and the limit exists. The limit $f \rightarrow 0$ with nonextremality condition ($f^{\prime} \neq 0$) gives the expression of $s_{ij}$ at the horizon of the nonextremal R.N. black hole,
\begin{equation}\label{shor}
\frac{s_{ij}}{2m^2}|_{Hor,Non-Ext}~=~1+\frac{\eta}{2} \Bigg[\frac{{\cal E}_{(i)}(r_H)}{{\cal E}_{(j)}(r_H)}\Bigg(1+\frac{L_{(j)}^2}{r_H^2}\Bigg)+\frac{{\cal E}_{(j)}(r_H)}{{\cal E}_{(i)}(r_H)}\Bigg(1+\frac{L_{(i)}^2}{r_H^2}\Bigg)-2\frac{L_{(i)}L_{(j)}}{r_H^2}\Bigg]
\end{equation}
with $r_H=M+\sqrt{M^2-Q^2}$ \& ${\cal E}_{(i)}(r_H)=E_{(i)}-\frac{e_{(i)}Q}{r_H}$. For $L_{(1)}=L_{(2)}=0$ (radial motion), $s_{12}$ reduces to the expression given in \cite{RN}.

If we take the extremal limit ($f^{\prime} \rightarrow 0$) after taking the horizon limit ($f \rightarrow 0$) the six kinematical invariants are found to be same as \eqref{shor} but with $r_H=M=Q$.

Both for nonextremal and extremal black holes, $s_{ij}$ grows without any bound for  a sharp kinematical asymmetry like $|{\cal E}_{(i)}(r_H)|>>|{\cal E}_{(j)}(r_H)|$, for generic values of $E$ and $Q$.  Notice that, if ${\cal E}_{(j)}(r_H)=0$, which may happen {\it without} $E_{(i)}$ being too large or small, but by simply relating $E_{(i)}$ to $e_{(i)} Q$, for generic values of all parameters, some invariants may become unbounded above. This is the limit when the rest energy of any particle is exactly equal to the electrostatic potential energy at the horizon. Thus, by choosing a proper kinematical regime, $e_{(i)}=\frac{E_{(i)}r_H}{Q}$, some kinematical  invariants can be made unbounded above.

\section{Particle Scattering in Kerr spacetime}

Kerr spacetime does not have any electromagnetic potential; particles interact only gravitationally in this spacetime. So the analysis is valid for both neutral and charged particles.

We choose the mass scale here; all quantities are given in the $M=1$ unit. The metric of the Kerr manifold of spin parameter $a$ is
\begin{equation}\label{Kerr Metric}
ds^2=\bigg(1-\frac{2r}{\Sigma}\bigg)dt^2+\bigg(\frac{4ar\sin^2\theta}{\Sigma}\bigg)dtd\phi-\bigg(\frac{\Sigma}{\Delta}\bigg)dr^2-\Sigma d\theta ^2-\bigg(r^2+a^2+\frac{2a^2r\sin^2\theta}{\Sigma}\bigg)sin^2\theta d\phi^2
\end{equation}
where $\Delta=r^2+a^2-2r$ and $\Sigma =r^2+a^2\cos^2\theta $.

We restrict ourselves to the particles moving in the equatorial plane $(\theta = \frac{\pi}{2})$ only. Geodesic equations corresponding to Eq. \eqref{Kerr Metric} is exactly integrated to get the four momenta \cite{BSW}, \cite{Carter},
\begin{equation}\label{fPK}
\frac{dt}{d\tau}=-\frac{1}{r^2}\bigg[a(aE-L)+\frac{(r^2+a^2)T}{\Delta}\bigg], ~\frac{dr}{d\tau}=\frac{1}{r^2}\sqrt{T^2-\Delta (r^2+(L-aE)^2)}, ~\frac{d\theta}{d\tau}=0, ~\frac{d\phi}{d\tau}=-\frac{1}{r^2}\bigg[(aE-L)+\frac{aT}{\Delta}\bigg]
\end{equation}
where $T=E(r^2+a^2)-La$ and E, L are the geodesic constants- energy and angular momentum per unit test mass.

Like the spherically symmetric spacetime, all of the six scalars are not independent here. The stationarity and $\phi$ independence of the metric coefficient guarantees conservation of $E$ and $L$, {\it i.e}, $E_{(1)}+E_{(2)}=E_{(3)}+E_{(4)}$ and $L_{(1)}+L_{(2)}=L_{(3)}+L_{(4)}$ for the scattering $1+2 \rightarrow 3+4$. The collision is restricted in the equatorial plane ($\theta$ is $\frac{\pi}{2}$ for all of the particles). Together, only three among the six scalars are independent for the collision of the particles in the equatorial plane of the Kerr spacetime. 

At any point in the equatorial plane during the geodesic motion, the six scalars appear to be
\begin{equation}\label{ssk}
\frac{s_{ij}}{2m^2}=1+\frac{\eta}{r \left(a^2+(r-2) r\right)}\\(-\sqrt{a^2 \left(E_{(i)}^2 (r+2)-r\right)-4 a E_{(i)} L_{(i)}+r^2 \left(\left(E_{(i)}^2-1\right) r+2\right)-L_{(i)}^2 (r-2)} 
\end{equation}
\begin{center}
$\hspace{2.5cm}\sqrt{a^2 \left(E_{(j)}^2 (r+2)-r\right)-4 a E_{(j)} L_{(j)}+r^2 \left(\left(E_{(j)}^2-1\right) r+2\right)-L_{(j)}^2 (r-2)}+a^2 E_{(i)} E_{(j)} r+2 a^2 E_{(i)}E_{(j)}$
\end{center}
\begin{center}
$-2 a E_{(i)} L_{(j)}-2 a E_{(j)} L_{(i)}+E_{(i)} E_{(j)} r^3-L_{(i)} L_{(j)} r+2 L_{(i)} L_{(j)})$
\end{center} 
where the particles are assumed to have same mass $m$. At the horizon, $r=r_H=1+\sqrt{1-a^2}$ and Eq. (\ref{ssk}) is in zero by zero form. Assuming nonextremality of the Kerr manifold $(a \neq 0)$, the horizon limit reduces Eq. (\ref{ssk}) to ($\Omega_H=\frac{a}{r_H^2+a^2}$ and ${\cal E}_{(i)}=E_{(i)}-\Omega_HL_{(i)}$; the effective energy of the particle in the Kerr spacetime)
\begin{equation}\label{skh}
\frac{s_{ij}}{2m^2}|_{Hor,Non-Ext}=2+\frac{\eta}{2} \bigg[\frac{{\cal E}_{(i)}}{{\cal E}_{(j)}}\bigg(1+\frac{L_{(j)}^2}{4}\bigg)+\frac{{\cal E}_{(j)}}{{\cal E}_{(i)}}\bigg(1+\frac{L_{(i)}^2}{4}\bigg)-\bigg(2+\frac{L_{(i)}L_{(j)}}{2}\bigg)\bigg]
\end{equation}
The expression for $s_{12}$ has already been derived in Ref. \cite{Harada1}. The expression \eqref{skh} is very much similar to that of the charged particle in Reissner-Nordstr\"om spacetime \eqref{shor} except by an additional factor of 2 in the last term of \eqref{skh}. This is due to the presence of the cross term in the metric of the Kerr spacetime. 

For an extremal Kerr manifold ($a=1$), Eq. (\ref{skh}) is also in zero by zero form. So the introduction of extremality after taking the horizon limit gives the possibility to take another limit $a \rightarrow 1$ which gives the same expression as Eq. (\ref{skh}) but now $r_H=1$ and $a=1$.

Like the charged particle in Reissner Nordstr\"om, here also expressions (\ref{skh}) suggest kinematical unboundedness of some invariants for the sharp kinematical asymmetry $|{\cal E}_{(i)}|>>|{\cal E}_{(j)}|$, for generic values of the parameters of the particles and spacetime, viz., $E~,~\Omega~,L$. This can be easily achieved, for instance, by setting ${\cal E}_{(j)}=0$ {\it i.e.}, by tuning the angular momentum of the particle, $L_{(j)}=\frac{E_{(j)}}{\Omega_H}$, without any parameter actually becoming unnaturally high. It is the limit when the energy of the particle becomes equal to the frame dragging rotational energy of the particle.

\section{Charged Particle Scattering in Kerr-Newman Black Hole Spacetime}

The metric of the Kerr-Newman spacetime is the same as Kerr spacetime \ref{Kerr Metric} except the fact that now $\Delta=r^2+a^2-2r$ is modified to $r^2+a^2-2r+Q^2$ ($Q$ is the charge of the spacetime and the mass scale is chosen). First, note that we are interested in the equatorial scattering only and the metric has stationarity and polar symmetry. Like before, this reduces the number of independent Mandelstam invariants to {\it three}.

For the neutral particle collision (having the same mass $m$) at the outer horizon of the Kerr-Newman black hole, the invariants are given by \eqref{skh} with the same divergence condition as mentioned in the last section except for now the radius of the outer horizon gets modified to $r_H=1+\sqrt{1-Q^2-a^2}$ and the condition for the extremality becomes $Q^2+a^2=1$.
 
In the Reissner-Nordstr\"om background, the neutral to charged particle transition is realized via the replacement $E_{(i)} \rightarrow {\cal E}_{(i)}$ as evident from the expressions of the invariants \eqref{shorizon} and \eqref{shor}.  Following the same logic, the expression of the kinematical scalars for the charged particle scattering in the Kerr-Newman spacetime is obtained replacing ${\cal E}_{(i)}$ by ${\cal E}_{(i)} \rightarrow {\cal E}_{(i)}-\frac{e_{(i)}Q}{r}$ in \eqref{skh} and [$e_{(i)}$=charge of the $i^{th}$ particle]. But apart from this, Kerr Newman spacetime possesses a magnetic vector potential $A_{\phi}$ \cite{Hejda-ref}. The magnetic vector potential at the horizon of the Kerr Newman black hole is calculated to be $A_{\phi}=\frac{Qa}{r_H}$ \cite{Kerr-Bardeen}. Due to the presence of a magnetic vector potential the angular momentum of the particle changes as
$L_{(i)}-\frac{e_{(i)}}{m}A_{\phi}$-    
\begin{equation}\label{Kerr Newman}
\frac{s_{ij}}{2m^2}=2+\frac{\eta}{2} \bigg[\frac{{\cal E}_{(i)}}{{\cal E}_{(j)}}\bigg(1+\frac{(L_{(j)}-\frac{e_{(j)}}{m}A_{\phi})^2}{4}\bigg)+\frac{{\cal E}_{(j)}}{{\cal E}_{(i)}}\bigg(1+\frac{(L_{(i)}-\frac{e_{(i)}}{m}A_{\phi})^2}{4}\bigg)-\bigg(2+\frac{(L_{(i)}-\frac{e_{(i)}}{m}A_{\phi})(L_{(j)}-\frac{e_{(j)}}{m}A_{\phi})}{2}\bigg)\bigg]
\end{equation}
${\cal E}_{(i)} = E_{(i)}- \Omega_H L_{(i)} - \frac{e_{(i)}Q}{r_H}$ and $\Omega_H=\frac{a}{r_H^2+a^2}$.  

For the extremal Kerr-Newman black hole, {\it six}  scalars are given by the same \eqref{Kerr Newman} with the constraint $Q^2+a^2=1$ and $r_H=1$.

Just as in every case considered before, the sharp kinematical asymmetry $|{\cal E}_{(j)}|>>|{\cal E}_{(i)}|$ leads to some invariants becoming unbounded above. A way to achieve this is to set ${\cal E}_{(i)}=0$ keeping ${\cal E}_{(j)}$ nonzero, for generic values of the particle and spacetime parameters. This implies a tuning of asymptotic energy of the particle $E_{(i)}=\Omega_H L_{(i)}-\frac{e_{(i)}Q}{r_H}$. 

\section{Conclusion}

The main aim of this paper has been to show that a total of six kinematical invariants are needed to characterize a collision of two particles kinematically, 
in a general curved spacetime. Isometries in the spacetime may reduce the number of independent invariants to less than six, depending on the nature of the 
spacetime. Only {\it three} scalars among {\it six} are independent for equatorial particle scattering in asymptotically flat Schwarzschild, Reissner-Nordstr\"om, Kerr and Kerr-Newman stationary black hole backgrounds. The kinematical scalars are functions of the effective energies of the colliding particles at the horizon of the above mentioned black holes, expressed in terms of a specific combination of parameters of the particle trajectory and the black hole spacetime. A sharp asymmetry between the effective energies between any pair of particles results in unboundedness above of the corresponding invariant. The center of mass energy in the CM frame of the particles 1 and 2 ($s_{12}$ in our notation) has been derived in all kinds of asymptotically flat black holes (perhaps in a different notation) in \cite{BSW}, \cite{RN}, \cite{Kerr-Newman}, \cite{Kerr-Newman1}, \cite{Hejda-ref}, and in agreement with our results. As expected from the previously found center of mass-energy expression, for charged particle scattering in Reissner-Nordstr\"om and Kerr-Newman horizon and for any general particle collision in Kerr horizon, the divergence can be achieved either by tuning the charge [$e_{(i)}=\frac{E_{(i)}r_H}{Q}$ in case of RN and $E_{(i)}=\Omega_H L_{(i)}+\frac{e_{(i)}Q}{r_H}$ in case of KN] or by tuning the angular momentum [$L_{(i)}=\frac{E_{(i)}}{\Omega_H}$ in case of Kerr], for generic values of the spacetime and particle parameters.

Throughout the paper, we assume that the colliding and the produced particles have the same mass. This simplifies our calculation but the main feature: i.e., the kinematical unboundedness of the necessary parameters is independent of this approximation. Some physicists have already relaxed this approximation for the center of mass-energy calculation \cite{Kerr-Newman}, \cite{Kerr-Newman1}, \cite{Hejda-ref}.

In flat spacetime, a large $s$ with a small $t$ causes the production of a shock wave \cite{Hooft}. One natural question arises- is there any analog of this effect in a generally curved region. To resolve this question one needs to investigate the dynamics of the two-particle scattering in various kinematical regimes. The effect analogous to the shock production is expected in these scenarios.

The result presented here is generic and ready for the application in astrophysical black holes. The unboundedness in the kinematical invariants can be interpreted as the high center of mass energy and/or high momentum transfer among the particles keeping the quantities measured from infinity (energy or angular momentum or charge) finite. It is explicitly discussed in \cite{Miyamoto} for the center of mass energy and can be realized for the other scalars also as the kinematical unboundedness of the Mandelstam like scalars do not require the divergence of the parameters tuned from asymptotically flat spacetime. This opens up the possibility to astrophysically check the physics beyond standard model. 

\section{Acknowledgement}

The authors gratefully acknowledge F Hejda for extremely valuable correspondence, especially for his remarks leading us to the correct results on the invariants for a Kerr-Newman black hole. SKR acknowledges Pratyusava Baral and Pronobesh Maity of for the helpful discussion. R. Koley acknowledges WBDHESTBT research grant.

\end{document}